\documentclass[aps,twocolumn,prd,showpacs,nofootinbib]{revtex4}
\usepackage{amsmath}
\usepackage{graphicx}
\usepackage{dcolumn}
\usepackage{bm}
\usepackage{amssymb}
\usepackage{latexsym}

\def\be{\begin{equation}}
\def\ee{\end{equation}}
\def\ba{\begin{eqnarray}}
\def\ea{\end{eqnarray}}

\begin{document}

\title{Non-Gaussianity in Island Cosmology  }

\author{Yun-Song Piao}

\affiliation{College of Physical Sciences, Graduate School of
Chinese Academy of Sciences, Beijing 100049, China}

\begin{abstract}

In this paper we fully calculate the non-Gaussianity of primordial
curvature perturbation of island universe by using the second
order perturbation equation. We find that
for the spectral index $n_s\simeq 0.96$, which is favored by
current observations, the non-Gaussianity level $f_{NL}$ seen in
island will generally lie between 30 $\sim$ 60, which may be
tested by the coming observations.
In the landscape,
the island universe is one of anthropically acceptable
cosmological histories. Thus the results obtained in some sense
means the coming observations, especially the measurement of
non-Gaussianity, will be significant to make clear how our
position in the landscape is populated.


\end{abstract}


\maketitle

The vacua in the landscape will be populated during eternal
inflation, see e.g. Refs. \cite{S06, G07, V06} for recent reviews.
In an anthropical viewpoint, how a vacuum like ours is populated
may be more crucial, since the history of populating determines
our observations. Recently, it has been argued that the island
cosmology in the landscape can be consistent with our real world
\cite{Piao0706b}, see earlier Refs.\cite{DV, Piao0506} for
discussions based on the background with the cosmological constant
observed.
The large fluctuations with the null energy condition violation
can stride over the barrier between vacua, and straightly create
some regions full with radiation, i.e. islands, in new or baby
vacua. These islands will evolve with the standard cosmology, some
of which under certain conditions may correspond to our observable
universe, see Ref.\cite{Piao0712} for details. In usual viewpoint,
in order to have an universe like ours in the landscape, the slow
roll inflation with enough period is generally required
\cite{FKMS}. This only can be implemented by a potential with a
long plain above corresponding minimum, which obviously means a
fine tuning, since the regions with such potentials are generally
expected to be quite rare in a random
landscape. 
While the island can actually emerge for any potential,
independent of whether the potential has a long plain. Thus in
principle as long as we can wait, the islands of observable
universes will be able to appear in any corner of landscape.

The island universe model brings a distinct anthropically
acceptable cosmological histories. Thus it is quite interesting to
ask how we can know whether we live in an island or in a reheating
region after slow roll inflation, which might be significant to
understand why and how our vacuum in the landscape is selected. In
principle, this can be judged by the observations of primordial
perturbations. However, in the level of first order scalar
perturbation, the island universe is actually degenerated with the
slow roll inflation, which in some sense is a reflection of
duality between their background evolutions, i.e. between the slow
expansion \cite{PZ} and the nearly exponent expansion, see Ref.
\cite{Piao0705, Piao0712} for details. Thus in principle it is
hardly possible to distinguish them by the spectrum index and
amplitude of curvature perturbation. However, recently, it has
been found that the non-Gaussianity of perturbation in island
cosmology is generally large \cite{Piao0712}, while that predicted
by the simple slow roll inflation model is quite small. Thus in
this sense the non-Gaussianity might be a powerful discriminator.

The current bound placed by WMAP5 is $-9< f_{NL} <111$
\cite{WMAP5}, which seems slightly prefer a net positive $f_{NL}$,
though $f_{NL}=0$ is still at $95\%$ confidence,
The analysis of large scale structure
combined with WMAP5 gave further limit $-1<f_{NL}< 70$
\cite{SHSHP}.
Further, the future Planck satellite
will be expected to give $\Delta f_{NL}\sim 5$ \cite{CSS},
These valuable observations are placing the island
universe to an interesting and tested regime. In Ref.
\cite{Piao0712}, the non-Gaussianity is rough estimated in term of
three point function, which is only determined by cubic
interaction term of field. However, this neglects other sources
for non-Gaussianity. Here the curvature perturbation is actually
induced by the entropy perturbation, thus the nonlinear relation
between the curvature perturbation and the entropy perturbation
can also contribute the non-Gaussianity. This is reflected in the
second order perturbation equation correlating both. In the
landscape, the island universe is one of anthropically acceptable
cosmological histories. It seems that the coming observations,
especially the measurement of non-Gaussianity, have had the
ability to identify cosmological history in which we live, and
thus show how our position in the landscape is populated. Thus in
order to have a definite prediction tested by coming and precise
observations, a full study for the non-Gaussianity of island
universe is obviously urgently required. This will be done in this
paper by applying the second order perturbation equation.

When the island emerges, the change of local background may be
depicted by $\epsilon\ll -1$, which is determined by the evolution
of local Hubble parameter `$h$', where the `local' means that the
quantities, such as the scale factor `$a$' and `$h$', only denote
the values of the null energy condition violating region.
$\epsilon\ll -1$ means the energy density of local emerging island
is rapidly increased. In order to phenomenologically describe and
simulate this behavior, we appeal the field same with the normal
scalar fields but with the minus sign in their kinetic terms,
which is usually called as ghost field. The evolution of such
ghost field is climbing up along its potential, and the steeper
its potential is, the faster it climbs, which is determined by the
property of this kind of field, e.g. Ref. \cite{GPZ}. Thus in Ref.
\cite{Piao0506}, it has been argued that such field can be
suitable for depicting the emergence of island. In the scenario of
island universe, as depicted in details in Refs. \cite{Piao0706b,
Piao0712}, initially the background is dS's, and then in some
regions the islands emerge, in which the local background
experiences a jump. There actually are not ghost fields presented
in entire scenario, since this phenomena is quantum. The ghost
field is that we introduce artificially, since we found that in
classical sense it can describe the evolution of emerging island
well, which make us be able to semiclassically explore the island
universe model and its possible predictions. In this sense, the
ghost field introduced only serves the evolution of background, by
which we can do some analytical and numerical calculations for
primordial perturbations. Further, for this purpose, this
introduced field should be required to satisfy some conditions
which assures the scenario of island universe not changed, for
example, it is not expected to participate in other quantum
processes.

We assume that $\epsilon $ is constant during the emergence of
island for simplicity. Thus
we can have the scale factor \be a\sim {1\over (-t)^{{1\over
|\epsilon|}}}\sim h^{1\over |\epsilon |}, \label{a}\ee which is
nearly unchanged since $|\epsilon|\gg 1$, which in some sense is
also why we call such a fluctuation as an emergent island, see
Fig.1 in Ref. \cite{Piao0512}. Thus the efolding number of mode
with some scale $\sim 1/k$ leaveing the horizon before the
thermalization can be written as ${\cal N} \simeq \ln{({h_e\over
h_i})}$ \cite{Piao0506}, where the subscript `$i$' and `$e$'
denote the initial and end values of relevant quantities,
respectively.
The observable cosmology requires ${\cal N}\sim 50$.
Thus in order to have an enough efolding number, an enough low
scale of parent vacuum should be selected.

The emergence of island in the landscape will generally involve
the upward fluctuations of a number of fields, or moduli. Thus it
is inevitable that there are entropy perturbations, which can
source the curvature perturbation. The method that we calculate
the curvature perturbation is similar to that applied in ekpyrotic
model \cite{LMTS, BKO}, see also \cite{KW} and earlier Refs.
\cite{NR, DFB}. The calculation of the non-Gaussianity is similar
to that implemented in Refs. \cite{LS1, LS2, KMVW, BKO1}. The
difference lies in the character of the fields used. Here as has
been mentioned, the normal scalar fields but with the minus sign
in their kinetic terms are used.
Thus compared to the corresponding equations for perturbations of
normal scalar fields, there will be some slight discrepancies in
relevant perturbation equations, i.e. difference of sign before
some terms, which, however, will lead to distinct results.

In principle, for both such fields, the rotation in field space
can be made, which decomposes fields into the field $\varphi$
along the motion direction in field space, and the field $s$
orthogonal to the motion direction \cite{GWBM}. In this case the
evolution of background will only determined by $\varphi$, whose
potential is only relevant with the background parameter
$|\epsilon|$, while $s$ will only contribute the entropy
perturbation, see Ref. \cite{Piao0712} for details. Here $v_k=
a\delta s_k$ is set for our convenience, and thus
$v_k^{(i)}=a\delta s_k^{(i)}$, where the superscript denotes the
$i$th order perturbation. Hereafter, we will study the equations
of perturbations with this replacement. The equation of first
order entropy perturbation and the detailed analysis of solutions
have been presented in Ref.\cite{Piao0705, Piao0712}, which thus
will be neglected here.
In term of Ref. \cite{Piao0712}, the spectrum index of $\delta s$
field is \be n_{\delta s}-1 \simeq {2\over \epsilon},
\label{ns}\ee which means that the spectrum of entropy
perturbation is nearly scale invariant with a slightly red tilt,
since $\epsilon\ll -1$. Here we have assumed that usual quantum
field theory can be applied even for such ghost fields
\footnote{Here we need to a normal quantization condition, like
usual field theory, to set initial conditions for primordial
perturbation, which seems contradict with that of ghost field.
However, this might be justified as follows. Initially the
background is dS's, in which there are not ghost fields, thus in
principle the normal quantization condition of usual field theory
can be applied. Then the island emerges, the local background
enters into a null energy violating evolution, which the ghost
field is introduced to describe. Thus the primordial perturbation
induced by such fields must have a normal quantization condition
as its initial condition, or it can not be matched to that of
initial dS background. }. The amplitude of perturbation spectrum
is \be {\cal P}^{1/2}_{\delta s}= k^{3/2}
\left|{v_k^{(1)}(\eta_e)\over a}\right| \simeq {1\over
\sqrt{2}a(-\eta_e)}, \label{ps}\ee
which is calculated at the end time $\eta_e$ of null energy
violating evolution, i.e. the emergence of island, since the
amplitude of perturbation on super horizon scale is increased all
along up to the end \cite{Piao0712}, where $\eta$ is conformal
time. Noting $a$ is nearly unchanged, which can be given from
Eq.(\ref{a}) since $|\epsilon|\gg 1$ and is actually a reflection
that the island is emerging very quickly, we have $a\eta\simeq t$,
thus the amplitude of spectrum can be rewritten as ${\cal
P}_{\delta s}\simeq {1\over 2(-t_e)^2}$. We can see that these
results are only determined by the evolution of background during
the emergence of island, but not dependent of other details.

The entropy perturbation can source the curvature perturbation by
${\dot {\cal R}}^{(1)}\simeq {2h{\dot\theta}\over
{\dot\varphi}}\delta s^{(1)}$ \cite{GWBM}. Thus if
${\dot\theta}=0$, i.e. the motion in field space is a straight
line, the entropy perturbation will not couple to the curvature
perturbation.
However, when there is a sharp change of direction of field
motion, ${\dot\theta}$ must be not equal to $0$, in this case
$\dot {\cal R}^{(1)}$ will inevitably obtain a corresponding
change induced by $\delta s$. We take the rapid transition
approximation \footnote{Here, during the null energy violating
evolution, i.e. the emergence of island, there is ${\dot\theta}=0$
till the end time, however, around the end time ${\dot\theta}$
must deviate from 0, thus in this sense this corresponds to a
rapid transition for $\theta$. In general, the period that $\dot
\theta$ deviates from 0 is far shorter than that of
${\dot\theta}=0$, which is the meaning of rapid transition
approximation.
Noting the approximation used here is similar to that used in
Refs. \cite{LMTS, BKO, KW, KMVW}, in e.g. \cite{BKO}, this
approximation is called as the rapid transition approximation,
thus here we follow this term. The null energy violating
transition means the total period of the null energy violating
evolution, i.e. the emergence of island, in which
${\dot\theta}=0$, while ${\dot\theta}\neq 0$ occurs only around
its end time. }, which means that all relevant quantities at split
second before the thermalization are nearly unchanged but only
$\theta$ changes from its initial fixed value $\theta=\theta_{*}$
to $\theta\simeq 0$, and thus have \be {\cal R}^{(1)} \simeq {2h_e
\theta_{*} \over {\dot\varphi}}\delta s^{(1)}, \label{xi}\ee which
leads that ${\cal R}^{(1)}$ acquires a jump induced by the entropy
perturbation $\delta s^{(1)}$ and thus inherits the nearly scale
invariant spectrum of $\delta s^{(1)}$ given by Eq.(\ref{ns}). We
can substitute Eq.(\ref{ps}) and ${h^2\over {\dot\varphi}^2}=
{4\pi\over |\epsilon|}$ into Eq.(\ref{xi}), and obtain the
resulting amplitude of curvature perturbation as ${\cal
P}_{(\delta s\rightarrow {\cal R})}
\simeq 16\theta_*^2\cdot |\epsilon| {h_e^2\over \pi}$, which is
approximately $ |\epsilon|h_e^2$. We can see that it and
Eq.(\ref{ns}) can be related to those of the usual slow roll
inflation by replacing $\epsilon$ as $-{1\over \epsilon}$, which
actually exactly gives the spectral index and amplitude of slow
roll inflation to the first order of slow roll parameters, noting
that this duality is valid not only for constant $|\epsilon|$
\cite{Piao0705} but also when $|\epsilon|$ is changed
\cite{Piao0712}.

The intrinsic non-Gaussianity in entropy perturbation can be
generated during the emergence of island. This can be obtained by
considering the motion equation of second order entropy
perturbation, which is, when ${\dot \theta}=0$, \be
v_k^{(2)\prime\prime} +\left(k^2-f(\eta)\right)
v_k^{(2)}+g(\eta)(v_k^{(1)})^2=0, \label{vk2}\ee where
$f(\eta)={a^{\prime\prime}\over a}+a^2{\cal V}_{(2)}$ and
$g(\eta)= -{a {\cal V}_{(3)}\over 2}$. The sign between both terms
of $f(\eta)$ is plus while the sign in $g(\eta)$ is minus, which
is just reverse with that of normal scalar field \cite{LV}. Here
${\cal V}_{(i)}$ denotes the $i$ times derivative for $s$, and
${\cal V}_{(2)} \simeq {2\over a^2\eta^2}$ and ${\cal
V}_{(3)}\simeq {8\alpha\sqrt{\pi }\over a^2\eta^2}\sqrt{
|\epsilon|}$ for $|\epsilon|\gg 1$, which can be obtained by
Eq.(5) in Ref. \cite{Piao0712}, where the constant $ \alpha \equiv
\sqrt{1/x}-\sqrt{x})$, and $\theta={\rm arctg}{(x)}$ is determined
by the cubic interaction of potential on $s$ field.
We only care the
solution at long wavelength. Thus taking $k\rightarrow 0$, we can
obtain \be v_k^{(2)}\simeq { \alpha
\sqrt{\pi|\epsilon|}(v_k^{(1)})^2 \over a}. \label{vk22}\ee Thus
we have $\delta s^{(2)}\simeq \alpha \sqrt{\pi|\epsilon|}(\delta
s^{(1)})^2$, since $v_k^{(i)}=a\delta s^{(i)}$.

The curvature perturbation induced by second order of entropy
perturbation can be given as \be {\dot {\cal R}}^{(2)}\simeq
{2h{\dot \theta}\over {\dot\varphi}}\delta s^{(2)}
-{h(4{\dot\theta}^2-{{\cal V}}_{(2)})\over {\dot \varphi}^2}
(\delta s^{(1)})^2 \label{calr2}\ee on large scale. The only
difference from Ref. \cite{LV} is here that before ${{\cal
V}}_{(2)}$ is the minus sign. The non-Gaussianity is generated
when modes are outside the horizon, thus here the non-Gaussianity
is expected to be local. The level of non-Gaussianity is usually
expressed in term of parameter $f_{ NL}$ defined in Refs.
\cite{KS, BCZ} \ba f_{ NL}& & =  -{5{\cal R}^{(2)}\over 3({\cal
R}^{(1)})^2} \simeq -{5\over 3({\cal R}^{(1)})^2}\nonumber\\ & &
\int \left({2h{\dot \theta}\over {\dot\varphi}}\delta s^{(2)}
-{h(4{\dot\theta}^2-{{\cal V}}_{(2)})\over {\dot \varphi}^2}
(\delta s^{(1)})^2 \right)dt, \label{fnl0}\ea where
Eq.(\ref{calr2}) has been applied.

\begin{figure}[t]
\begin{center}
\includegraphics[width=7cm]{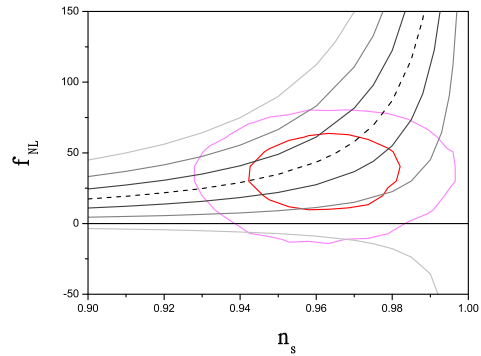}
\caption{
The $f_{NL}-n_s$ plane, in which the solid lines from top to down
correspond to $\theta_*= 0.7,0.8,0.9,1.1,1.2,1.3$, respectively.
The dashed line is that of $\theta_*=1.0$. The $1\sigma$ and
$2\sigma$ contours on $f_{NL}-n_s$ is plotted by using the data in
Ref. \cite{SHSHP}. We can see that for $\theta_*\simeq 1.0$,
$f_{NL}\simeq 30\sim 60$ is definitely predicted by the current
observations for $n_s$. }
\end{center}
\end{figure}

The terms in Eq.(\ref{fnl0}), proportional to ${\dot \theta}$, are
not 0 only at split second before the thermalization. Thus the
rapid transition approximation can be applied in the calculations.
The first term corresponds to the intrinsic non-Gaussiantity of
$\delta s$. This can be inherited by the curvature perturbation,
which is \be -{5\over 3({\cal R}^{(1)})^2}\int
{2h{\dot\theta}\over {\dot\varphi}}\delta s^{(2)} dt \simeq
-{5\alpha\over 12 \theta_*}|\epsilon|, \label{f1}\ee where
Eqs.(\ref{xi}) and (\ref{vk22}) have been used. This result in
fact equals to that calculated by using the three point function
\cite{Piao0712}. The second term in Eq.(\ref{fnl0}) corresponds to
the nonlinear correction for the linear relation between $\cal R$
and $\delta s$. It will also contribute the non-Gaussiantity of
the curvature perturbation, which is
\be {5\over 3({\cal R}^{(1)})^2} \int {h(4{\dot\theta}^2-{{\cal
V}}_{(2)})\over {\dot \varphi}^2} (\delta s^{(1)})^2 dt   \simeq
{5\over 6\theta_*}|\epsilon|, \label{f2}\ee where ${{\cal
V}}_{(2)}\simeq {2\over a^2\eta^2}\simeq {2\over |t|^2}$ for
$|\epsilon|\gg 1$, and also we set ${\dot\theta}\simeq {1\over
|t|}$ for calculation. The latter means the period $\Delta t_*$ of
change of $\theta$ can be deduced from $\int{\dot \theta}dt \simeq
1$. Thus we have $\Delta t_*\simeq |t_e|\simeq {1\over |\epsilon|
h_e}$, noting that $t$ is negative. While the total time that the
emergence of island lasts is $ T
\simeq {1\over |\epsilon | h_i}$ \cite{Piao0506, Piao0712}, which
is far shorter than one Hubble time since $|\epsilon|\gg 1$ and
thus is consistent with the claim that the emergence of the island
is a quantum fluctuation in the corresponding dS background. The
enough efolding number require $h_e/h_i\gtrsim e^{50}$, thus we
have $ T\simeq \Delta t_*e^{50}$, i.e. the period of change of
$\theta$ is far less than the time the emergence of island lasts,
This is consistent with the rapid transition approximation used.

The term in Eq.(\ref{calr2}), proportional to ${{\cal V}}_{(2)}$,
is not relevant with $\dot \theta$. Thus there exists a nonlinear
dependence of $\cal R$ to $\delta s$ during the entire evolutive
period of fluctuation. In this case this term will contributes an
integrated non-Gaussianity. When ${\dot\theta}=0$,
Eq.(\ref{calr2}) becomes ${\dot {\cal R}}^{(2)} = {h{{\cal
V}}_{(2)}(\delta s^{(1)})^2\over {\dot \varphi}^2}$.
Then we make the integral for this equation, and can obtain the
relation of $ (\delta s^{(1)})^2\sim -{{\cal R}^{(2)}}$, noting
that here Eq.(\ref{ps}) need to be used. Thus the contribution of
this integral effect for non-Gaussianity can be written as \be
-{5{\cal R}^{(2)}\over 3({\cal R}^{(1)})^2}\simeq {5\over
12\theta_*^2}|\epsilon|, \label{f3}\ee which is inverse to
$\theta_*^2$, not like Eqs.(\ref{f1}) and (\ref{f2}). When
$\theta_*\ll 1$, this term will make $f_{ NL}$ very large.

Thus the total non-Gaussianity of the
curvature perturbation is 
\be f_{NL} \cong {5\left(-\alpha\theta_*+2\theta_*+1\right)\over
12\theta^2_*}|\epsilon|, \label{ftot}\ee which is the sum of the
results given in Eqs.(\ref{f1}), (\ref{f2}) and (\ref{f3}). We can
see that in general the non-Gaussianity in island cosmology is
large, since
$|\epsilon|$ is large. However,
since here $\alpha$ is also the function of $\theta_*$,
where $\theta_*$ takes its value between 0 and $\pi/2$, thus for a
fixed $|\epsilon|$, the value of $f_{NL} $ may be larger or
smaller dependent of $\theta_*$.
In general without any fine tuning, $\theta_*$ should be about 1.
For $\theta_*\simeq 1$, and $n_s\simeq 0.96$ meaning
$|\epsilon|\simeq 50$ from Eq.(\ref{ns}), we can have
$f_{NL}\simeq 43$, which is a preferred positive value by the
current observations. While a smaller $\theta_*$ means a larger
fine tuning, and also a larger $f_{NL}$, which is not favored. In
addition, in principle there can be an accident cancellation for
all $\theta_*$-dependent terms in Eq.(\ref{ftot}) for some value
of $\theta_*$, in this case $f_{NL}\simeq 0$. This value is about
1.26, beyond which $f_{NL}<0$.

We can obtain $f_{ NL}\sim {1/|n_s-1|}$ by combining
Eqs.(\ref{ns}) and (\ref{ftot}), which means that $f_{NL}\sim
{\cal O}(10)$ since the red shift $|n_s-1|>0.01$, and the redder
the spectrum is the smaller $f_{NL}$ is. The reason is that a
redder spectrum corresponds to a smaller $|\epsilon|$, thus
$f_{NL}$. This result is different from that in simple slow roll
inflation model, in which $f_{NL}$ is not inverse proportional to
$|n_s-1|$ like in island, but proportional to it, e.g. Ref.
\cite{M02}. This predestines that the non-Gaussianity in simple
slow roll inflation is quite small. We plot a $f_{NL}-n_s$ plane
in Fig.1 for further illustration. This figure can be
distinguished from that in ekpyrotic and cyclic model \cite{LS1,
LS2}, in which in principle the redder the spectrum is, the larger
the non-Gaussianity is, see also Fig.5 in Ref. \cite{SHSHP}.
Though it seems that there requires $|\epsilon|\gg 1$ both in our
model and in cyclic model, and the only difference is that
$\epsilon$ is negative in our model and positive in the latter, it
is this difference that leads that their behavior is distinctly
contrary in $f_{NL}-n_s$ plane. In cyclic model, the spectrum
index obtained is the same with that the island universe model.
However, since $\epsilon\gg 1$, when $\epsilon$ is constant, the
spectrum will be blue, which can be seen in Eq.(\ref{ns}). Thus to
have a red spectrum favored by the observations, the change of
$\epsilon$ must be considered. In this case, the spectrum index is
$n_s-1 \simeq {2\over \epsilon}-{d\ln{|\epsilon|}\over d{\cal
N}}$. The red spectrum requires ${d\ln{|\epsilon|}\over d{\cal
N}}>{2\over \epsilon}$. This may be implemented only by
introducing a larger $\epsilon$, since this can lead to a smaller
${2\over \epsilon}$. Thus in this case a redder spectrum
corresponds to a larger $f_{NL}$. In order to have an enough red
spectrum, for example $n_s\simeq 0.97$, $\epsilon$ must be large
and change with ${\cal N}$ more rapidly than $\epsilon\sim {\cal
N}$. However, in island universe, this is not necessary, since
$\epsilon\ll -1$, which assure that its spectrum is naturally red.
Including the change of $\epsilon$ dose not alter our result
qualitatively.

In Eq.(\ref{ftot}), $f_{NL}\sim |\epsilon|$ should be general,
since $|\epsilon|$ is only determined by the evolution of
background, which is independent of modeling. While the details of
modeling only change the factor between $|\epsilon|$. It is
inevitable that this factor is dependent of the parameters of
model. However, this dependence is actually not important for the
natural values of parameters of model, here it is obvious that the
resulting $f_{NL}$ is mainly determined by $|\epsilon|$. The
generalization of $f_{NL}\sim |\epsilon|$ can be also seen for
simple slow roll inflation, in which $f_{NL}\sim \epsilon$, e.g.
\cite{M02}. It can be noted that in ekpyrotic and cyclic model
\cite{LS1, LS2}, $f_{NL}\sim \sqrt{\epsilon}$. This is because
they required the entropy perturbation induces the curvature
perturbation occurs during the kinetic energy domination after
ekpyrotic phase. When it is required to occur during ekpyrotic
phase, the result will be same with $f_{NL}\sim |\epsilon|$.
However, in this case, as has been mentioned, in order to have a
red tilt spectrum, a larger $\epsilon$ must be introduced, which
will conflict with the bound for non-Gaussianity from current
observations. Thus in there this case is not adopted.


In summary, the non-Gaussianity of island universe model is
calculated fully by using the second order perturbation equation.
We found that for the best fit value $n_s\simeq 0.96$ given by the
current observations, without any fine tuning of relevant
parameter, $f_{NL}\simeq 43$, which is about between $30$ $\sim$
$60$ when the uncertainty for $n_s$ from WMAP5 is included.
In simple slow roll inflation model, the non-Gaussianity is
generally quite small. Thus in order to obtain a large positive
value, some special operations for perturbations or models must be
appealed, which leads that its prediction has certain randomicity.
Thus compared with the inflation, the distinct prediction of
island universe for the non-Gaussianity makes it be able to be
falsified definitely by coming observations. In this sense if the
cosmological dynamics is actually controlled by a landscape of
vacua, the results of coming observations, especially the
measurement of non-Gaussianity, will be significant to make clear
whether we live in an island or in a reheating region after slow
roll inflation, which will be significant to understand why and
how our position in the landscape is populated.

\textbf{Acknowledgments} We thank A. Slosar for sending us the
data in Ref. \cite{SHSHP}. This work is supported in part by NNSFC
under Grant No: 10775180, in part by the Scientific Research Fund
of GUCAS(NO.055101BM03), in part by CAS under Grant No:
KJCX3-SYW-N2.

\end{document}